\documentclass[conference]{IEEEtran}

\usepackage[T1]{fontenc} 
\IEEEoverridecommandlockouts

\IEEEoverridecommandlockouts
\usepackage{cite}
\usepackage{balance}
\usepackage{amsmath,amssymb,amsfonts}
\usepackage{algorithmic}
\usepackage{graphicx}
\usepackage{textcomp}
\usepackage{xcolor}
\def\BibTeX{{\rm B\kern-.05em{\sc i\kern-.025em b}\kern-.08em
    T\kern-.1667em\lower.7ex\hbox{E}\kern-.125emX}}
\begin{document}

\title{Hardware-Efficient Reinforcement-Learned Scheduling for Quantum LDPC Decoders\\

}

\author{\IEEEauthorblockN{Mohsen Moradi$^{1}$, Vahid Nourozi$^{2}$, Wei Tang$^{2}$, and David Mitchell$^{2}$}
\IEEEauthorblockA{1. School of Electrical, Computer and Energy Engineering, Arizona State University, Tempe, AZ, 85287, USA \\
2. Klipsch School of Electrical and Computer Engineering, New Mexico State University, Las Cruces, NM, 88003, USA\\
Corresponding author: \textit{dgmm@nmsu.edu}}

}

\maketitle

\begin{abstract}
Quantum low-density parity-check (QLDPC) codes are
promising for fault-tolerant quantum computing, but
their decoding algorithms present significant hardware
implementation challenges. Reinforcement-learned
sequential belief propagation (RL-S) improves the
error-correction performance of conventional belief
propagation (BP) through adaptive variable-node
scheduling. However, its heap-based priority management
introduces substantial hardware overhead. This paper
presents hardware implementations of BP, RL-S, and
bucket-based reinforcement-learned BP (BRL-S) for the
$[[288,12,18]]$ bivariate bicycle QLDPC code. The BRL-S
architectures employ quantized priorities and
first-in, first-out (FIFO) buckets to simplify scheduling.
Three configurations with two, four, and eight priority
levels are evaluated. FPGA synthesis results demonstrate
that BRL-S reduces lookup-table utilization by approximately
68--72\% compared with RL-S. ASIC synthesis further
shows reductions of 63--67\% in area and 64--69\%
in estimated power consumption. These results demonstrate
that bucket-based scheduling substantially improves
the hardware efficiency of reinforcement-learned
QLDPC decoders.
\end{abstract}


\section{Introduction}
\label{sec:introduction}

Quantum error correction (QEC) is essential for achieving
reliable quantum computation in the presence of physical
qubit errors. Quantum low-density parity-check (QLDPC)
codes have attracted increasing attention because of their
potential to provide effective error correction with
relatively low redundancy \cite{QLDPC,BBCode}.
However, decoding QLDPC codes requires processing
syndrome measurements through iterative algorithms,
creating substantial computational and hardware
implementation challenges. Practical QEC systems
therefore demand decoding architectures that provide
strong error-correction capability while maintaining
manageable hardware complexity and resource consumption.

Belief propagation (BP) is a widely studied decoding
algorithm for LDPC codes due to its iterative
message passing structure and suitability for hardware
implementation~\cite{valls2021syndrome,raveendran2023soft}. Conventional flooding BP updates
messages simultaneously across the Tanner graph according to a
predefined schedule, allowing substantial parallelism
in hardware. However, BP decoding of QLDPC codes is
particularly challenging because of short cycles,
degeneracy, and correlated error structures, which
can degrade convergence and error-correction
performance \cite{QuantumBP}.
{QLDPC hardware studies include min-sum and soft-syndrome
FPGA decoders~\cite{valls2021syndrome,raveendran2023soft},
check-agnosia architectures~\cite{ducrest2024checkagnosia},
and FPGA/ASIC implementations of BP with ordered-statistics
decoding for BB codes~\cite{bascones2025hardware}.
Other work explores sparse acceleration through
Vegapunk~\cite{zhou2025vegapunk} and reports preliminary
FPGA results for decoding correlated errors under
circuit-level noise~\cite{maan2026correlated}.}

{Sequential schedules can improve BP decoding
performance~\cite{Moradi2026Sequential}.}
Reinforcement-learned sequential BP
(RL-S) has recently been introduced to improve QLDPC
decoding by adaptively selecting variable nodes (VNs)
according to learned priorities \cite{RLS}.
Instead of following a fixed flooding schedule,
RL-S updates individual VNs in a priority driven
sequence, improving the effectiveness of message
propagation and reducing decoding errors.
However, conventional RL-S relies on a priority
queue, typically implemented using a max-heap,
to identify the highest-priority VN and maintain
the scheduling order. Heap-based scheduling requires
repeated comparisons and priority adjustments,
while storing fine-grained learned priorities
introduces additional memory and logic requirements.
These operations complicate efficient hardware
realization of RL-S decoders.

Bucket-based reinforcement-learned scheduling (BRL-S)
provides a hardware-friendly alternative by quantizing
learned priorities into a limited number of discrete
levels and organizing VNs into corresponding
first-in, first-out (FIFO) buckets \cite{BRL}.
The highest-priority nonempty bucket determines the
next VN to be updated, replacing heap-based priority
maintenance with simpler queue operations.
In this work, we develop synthesizable
register-transfer-level (RTL) architectures for
conventional BP, RL-S, and BRL-S decoders targeting
the $[[288,12,18]]$ bivariate bicycle QLDPC code.
Three BRL-S configurations with two, four, and eight
priority buckets are implemented to investigate
the effects of priority resolution on hardware
complexity.

The proposed BRL-S hardware architectures substantially
simplify the scheduling subsystem while retaining
the sequential BP message update
framework. Quantized priority representations
reduce the storage requirements associated with
learned scheduling, while FIFO-based bucket
management eliminates the need for heap reordering
operations. FPGA synthesis results show that BRL-S
implementations require approximately 20,000
lookup tables (LUTs) and 5,800 flip-flops (FFs),
compared with 69,737 LUTs and 10,536 FFs for RL-S.
ASIC synthesis further demonstrates substantial
reductions in standard cell area and estimated
power consumption. These results highlight the
potential of bucket-based scheduling to improve
the hardware efficiency of reinforcement-learned
QLDPC decoders.

The main contributions of this paper are
summarized as follows:
\begin{enumerate}
    \item \textbf{Hardware architecture development:}
    We develop synthesizable RTL implementations
    of flooding BP, RL-S, and BRL-S
    for the $[[288,12,18]]$ bivariate bicycle QLDPC
    code, enabling architecture level comparisons
    under a common decoding configuration;

    \item \textbf{Bucket-based scheduling implementation:}
    We implement quantized priority FIFO scheduling
    with two, four, and eight buckets, replacing
    the max-heap priority management of RL-S with
    simpler hardware structures while preserving
    the underlying sequential message update
    framework;

    \item \textbf{FPGA and ASIC evaluation:}
    We systematically evaluate the five decoder
    architectures using FPGA resource utilization
    and ASIC synthesis metrics, demonstrating
    the hardware cost reductions achieved through
    bucket-based scheduling and characterizing
    the trade-offs associated with different
    priority resolutions.
\end{enumerate}


\section{Background}
\label{sec:background}

\subsection{BP, RL-S, and BRL-S Decoding Algorithms}
\label{subsec:algorithms}

A Calderbank--Shor--Steane (CSS)~\cite{CalderbankShor,Steane} quantum low-density
parity-check (QLDPC) code is defined by two sparse
binary parity-check matrices, $\mathbf{H}_X$ and
$\mathbf{H}_Z$, satisfying
$\mathbf{H}_X\mathbf{H}_Z^T=\mathbf{0}$ over
$\mathbb{F}_2$. This work considers the
$[[288,12,18]]$ bivariate bicycle (BB) code~\cite{BBCode}, which
encodes 12 logical qubits into 288 physical qubits.
The principal symbols used in the decoding algorithms
are summarized in Table~\ref{tab:symbols}.

\begin{table}[t]
\centering
\caption{Symbol Explanation for QLDPC Decoding}
\label{tab:symbols}
\renewcommand{\arraystretch}{1.08}
\begin{tabular}{cl}
\hline
\textbf{Symbol} & \textbf{Description} \\
\hline
$n$ & Number of physical qubits \\
$\mathbf{H}_X,\mathbf{H}_Z$ &
    CSS parity-check matrices \\
$\mathbf{e}_X,\mathbf{e}_Z$ &
    Binary error component vectors \\
$\mathbf{s}_X,\mathbf{s}_Z$ &
    Measured syndrome vectors \\
$\hat{\mathbf{e}}_X,\hat{\mathbf{e}}_Z$ &
    Estimated error vectors \\
$\boldsymbol{\delta}_X,\boldsymbol{\delta}_Z$ &
    Residual syndrome vectors \\
$v_i$ & Variable node indexed by $i$ \\
$\sigma_i$ & Local residual syndrome state \\
$Q(\sigma_i,i)$ & Learned scheduling score \\
$P(\sigma_i,i)$ & Quantized scheduling priority \\
$K$ & Number of priority buckets \\
$\mathcal{R}$ & VNs remaining in current sweep \\
\hline
\end{tabular}
\end{table}

Under a depolarizing channel with physical error
probability $p$, each qubit experiences a Pauli error
$E_i\in\{I,X,Y,Z\}$ with probability
\begin{equation}
\Pr(E_i=q)=
\begin{cases}
1-p, & q=I,\\
p/3, & q\in\{X,Y,Z\}.
\end{cases}
\label{eq:depolarizing}
\end{equation}
The corresponding measured syndromes are
\begin{equation}
\mathbf{s}_X=\mathbf{H}_X\mathbf{e}_Z,\qquad
\mathbf{s}_Z=\mathbf{H}_Z\mathbf{e}_X,
\label{eq:syndrome}
\end{equation}
where all binary operations are performed over
$\mathbb{F}_2$. For the current error estimates,
the residual syndromes are defined as
\begin{equation}
\begin{aligned}
\boldsymbol{\delta}_X
&=\mathbf{s}_X+\mathbf{H}_X\hat{\mathbf{e}}_Z,\\
\boldsymbol{\delta}_Z
&=\mathbf{s}_Z+\mathbf{H}_Z\hat{\mathbf{e}}_X.
\end{aligned}
\label{eq:residual}
\end{equation}
A syndrome consistent estimate is obtained when
both residual syndrome vectors are zero.

\textit{1) Conventional Belief Propagation:}
BP performs iterative message
updates between variable nodes (VNs) and check nodes
(CNs) in the Tanner graph. The flooding schedule
updates all CN-to-VN messages and subsequently all
VN-to-CN messages within each global iteration.
The message computations follow the sum-product
algorithm (SPA), which combines incoming messages
to update the beliefs of the corresponding error
variables. Although flooding BP offers a regular
message update structure, short cycles and
degeneracy in QLDPC codes can degrade its
error correction performance.

\textit{2) Reinforcement-Learned Sequential BP:}
RL-S replaces the fixed flooding schedule with
adaptive VN selection based on a trained Q-table \cite{RLS}.
For each VN $v_i$, a local state $\sigma_i$ is
constructed from the residual syndrome bits of
its neighboring checks. Let $\mathcal{R}$ denote
the set of VNs that have not been updated in the
current sweep. RL-S selects the next VN according to
\begin{equation}
i^* =
\underset{i\in\mathcal{R}}{\arg\max}
\,Q(\sigma_i,i).
\label{eq:rl_selection}
\end{equation}
The selected VN is removed from $\mathcal{R}$
and updated using the sequential SPA procedure.
When the hard decision changes, the affected
residual syndrome states and scheduling priorities
are refreshed.
A max-heap maintains the priorities
of eligible VNs, avoiding a full search during
each selection. However, the fine-grained Q-table
and heap maintenance introduce additional
hardware storage and control complexity.

\begin{figure*}[ht!]
    \centering
    \includegraphics[width=\textwidth]{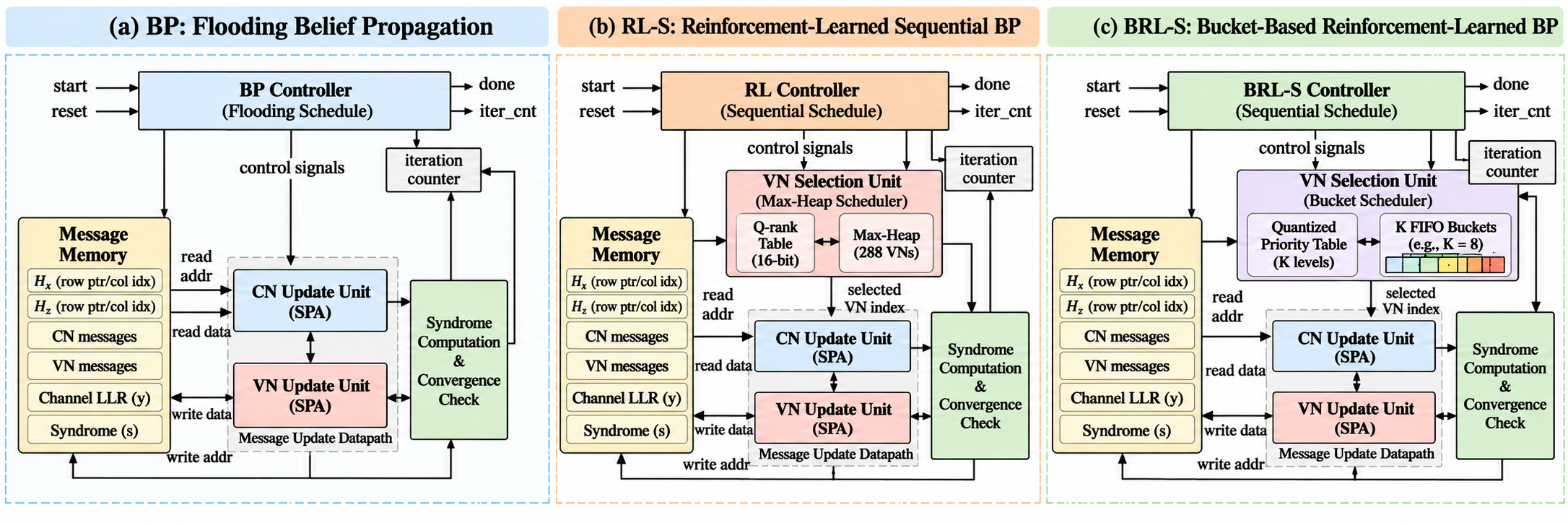}
    \vspace{-0.2in}
    \caption{Hardware architectures of three BP based
    decoders for the [[288,12,18]] bivariate bicycle quantum LDPC
    code: (a) conventional flooding BP, (b) reinforcement-learned
    sequential BP (RL-S) with max-heap priority scheduling, and
    (c) bucket-based reinforcement-learned BP (BRL-S) with
    quantized priorities and K FIFO buckets.
    RL-S and BRL-S employ sequential variable node updates,
    while BRL-S replaces heap-based scheduling with
    constant complexity bucket operations.}
    \label{fig:decoder_architecture}
    \vspace{-0.2in}
\end{figure*}

\textit{3) Bucket-Based Reinforcement-Learned BP:}
BRL-S simplifies RL-S scheduling by quantizing the
trained Q-values into $K$ discrete priority levels.
Let $Q_{\min}$ and $Q_{\max}$ denote the minimum
and maximum entries of the trained Q-table.
For $Q_{\max}>Q_{\min}$, the quantization interval
and priority assignment are
\begin{equation}
\begin{aligned}
\Delta_Q &= \frac{Q_{\max}-Q_{\min}}{K},\\
P(\sigma_i,i)
&=\min\left\{K-1,
\left\lfloor
\frac{Q(\sigma_i,i)-Q_{\min}}{\Delta_Q}
\right\rfloor\right\}.
\end{aligned}
\label{eq:priority_quantization}
\end{equation}
If all Q-values are equal, every priority is
assigned to level zero. The quantization is
performed offline, and only the resulting
priority table is stored in the decoder.

BRL-S maintains $K$ first-in, first-out (FIFO)
buckets, $\mathcal{B}_0,\ldots,\mathcal{B}_{K-1}$.
The next VN is selected from the highest priority
nonempty bucket as
\begin{equation}
\begin{aligned}
k^* &= \max\{k\>|\>\mathcal{B}_k\neq\emptyset\},\\
i^* &= \operatorname{head}(\mathcal{B}_{k^*}).
\end{aligned}
\label{eq:bucket_selection}
\end{equation}
After each VN update, only eligible VNs affected
by changes in the residual syndrome require
priority refresh. BRL-S retains the same sequential
SPA message update procedure as RL-S, while
replacing heap-based scheduling with FIFO bucket
management. This work evaluates $K=2$, $4$, and
$8$, denoted as BRL2, BRL4, and BRL8, respectively.

\subsection{Decoding Performance Comparison}
\label{subsec:decoding_performance}

\begin{figure}[t]
    \centering
    \includegraphics[width=\linewidth]
        {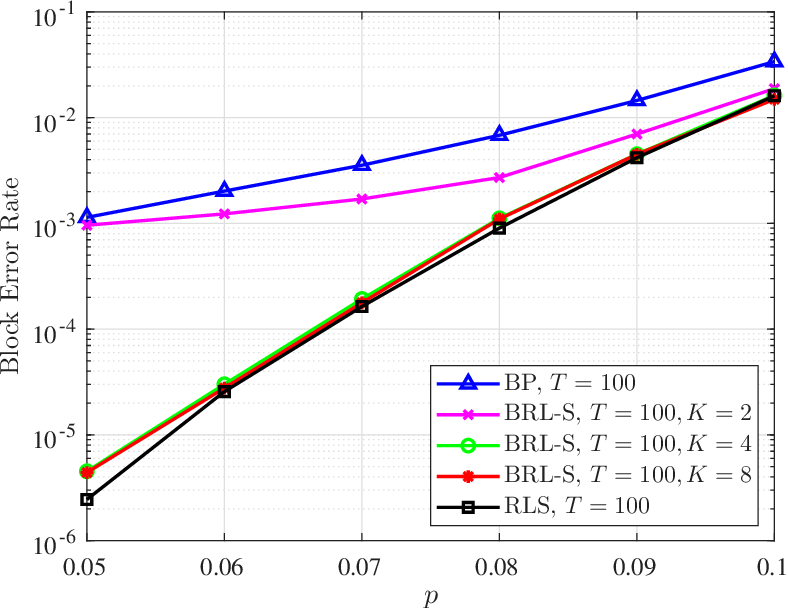}
        \vspace{-0.2in}
    \caption{BLER versus physical error probability $p$
for the $[[288,12,18]]$ BB code over the depolarizing
channel. Flooding BP, RL-S, and BRL-S with $K=2$,
$4$, and $8$ priority levels are compared, all with
a maximum of $T=100$ decoding iterations.}
    \label{fig:bler}
    \vspace{-0.2in}
\end{figure}

Fig.~\ref{fig:bler} compares the block error rate
(BLER) of conventional BP, RL-S, and BRL-S with
$K=2$, $4$, and $8$ priority levels for the
$[[288,12,18]]$ BB code over the depolarizing
channel. 
All decoders use a maximum of $T=100$ decoding
iterations. For sequential decoders, one iteration corresponds to one sweep of up to 288 variable node updates.
RL-S and all three BRL-S
configurations achieve substantially lower BLER
than conventional BP across the evaluated physical
error probabilities. In particular, BRL4 and BRL8
exhibit decoding performance close to that of RL-S,
indicating that a limited number of quantized
priority levels is sufficient to preserve the
benefits of reinforcement-learned scheduling.
BRL2 exhibits a modest performance degradation,
particularly at lower physical error probabilities.
These results demonstrate that BRL-S can maintain
competitive error correction performance with
reduced scheduling complexity, motivating the
hardware implementation and resource evaluation
presented in Section~\ref{sec:hardware}.

%
%


\section{Hardware Implementation}
\label{sec:hardware}

\subsection{Decoder Hardware Architectures}
\label{subsec:architecture}

Fig.~\ref{fig:decoder_architecture} illustrates the
hardware architectures of the conventional BP,
RL-S, and BRL-S decoders. All three architectures
target the $[[288,12,18]]$ BB code and contain
control logic, message storage, SPA computation,
and syndrome evaluation. Their primary architectural
difference lies in the scheduling and management
of VN updates.

\textit{1) Flooding BP Architecture:}
As shown in Fig.~\ref{fig:decoder_architecture}(a),
the conventional BP decoder employs a flooding
schedule controlled by the BP controller.
The message storage provides the parity-check
matrix connectivity, channel information,
and intermediate messages required by the
CN and VN update units. The SPA datapath
performs iterative message computations,
while the syndrome evaluation logic determines
whether the current error estimate satisfies
the measured syndrome. Unlike RL-S and BRL-S,
the flooding architecture does not require
learned priority storage or a VN selection unit.

\textit{2) RL-S Architecture:}
Fig.~\ref{fig:decoder_architecture}(b) shows
the RL-S decoder, which introduces a learned
priority scheduler to determine the VN update
order. The scheduler contains a 16-bit Q-rank
table and a max-heap storing the priorities
of up to 288 VNs. At each scheduling step,
the highest priority eligible VN is selected,
and its index is supplied to the message update
datapath. The decoder then performs the
corresponding CN and VN message computations
and updates the hard decision. When the
residual syndrome changes, the priorities
of affected eligible VNs are refreshed,
requiring heap maintenance operations.
The resulting architecture implements
sequential VN updates with adaptive scheduling.

\textit{3) BRL-S Architecture:}
Fig.~\ref{fig:decoder_architecture}(c) presents
the BRL-S decoder, which replaces the max-heap
scheduler with a quantized priority table
and $K$ FIFO buckets. The three configurations employ 1-bit, 2-bit, and 3-bit priority representations for $K=2$,  $4$, and $8$, respectively. Each bucket
maintains the eligible VNs assigned to
its priority level. A priority encoder
identifies the highest priority nonempty
bucket, and the VN at its head is selected
for processing. When an eligible VN's
priority changes, it is relocated to the
corresponding bucket. The queue structure
uses head and tail pointers together with
linked list pointers for VN management.

\begin{table}[t]
\centering
\caption{FPGA Resource Utilization Comparison of LDPC Decoders}
\vspace{-0.1in}
\label{tab:fpga_results}
\renewcommand{\arraystretch}{1.1}
\setlength{\tabcolsep}{3.5pt}
\begin{tabular}{llrrr}
\hline
\textbf{Method} &
\textbf{Code Type} &
\textbf{Length} &
\textbf{LUTs} &
\textbf{FFs} \\
\hline
Andrade et al. \cite{Andrade2015}
    & Non-binary LDPC & 384 & 350,892 & 294,576 \\
Ye et al. \cite{Ye2025ISCAS}
    & QKD LDPC & 100 & 379,121 & 39,743 \\
\hline
BP (This work)
    & BB QLDPC & 288 & 100,267 & 71,995 \\
RL-S (This work)
    & BB QLDPC & 288 & 69,737 & 10,536 \\
BRL2 (This work)
    & BB QLDPC & 288 & 19,873 & 5,784 \\
BRL4 (This work)
    & BB QLDPC & 288 & 22,078 & 5,790 \\
BRL8 (This work)
    & BB QLDPC & 288 & 20,505 & 5,865 \\
\hline
\end{tabular}
\vspace{-0.1in}
\end{table}

\begin{table}[h!]
\centering
\caption{ASIC Synthesis Results of BB288 QLDPC Decoders}
\vspace{-0.1in}
\label{tab:asic_comparison}
\footnotesize
\setlength{\tabcolsep}{5pt}
\begin{tabular}{lrrr}
\hline
\textbf{Decoder} &
\textbf{Area} &
\textbf{Power} &
\textbf{Clock Period} \\
& \textbf{(mm$^2$)} &
\textbf{(mW)} &
\textbf{(ns)} \\
\hline
BP   & 1.0710 & 15.243 & 11.414 \\
RL-S & 0.4564 & 3.766  & 2.589  \\
BRL2 & 0.1485 & 1.175  & 2.537  \\
BRL4 & 0.1591 & 1.266  & 2.513  \\
BRL8 & 0.1712 & 1.366  & 2.548  \\
\hline
\end{tabular}
\vspace{-0.1in}
\end{table}

The RL-S and BRL-S implementations employ
the same sequential SPA computation framework,
including CN message updates, VN belief
calculations, hard decisions, and residual
syndrome maintenance. Consequently, their
hardware comparison primarily reflects
the implementation cost of the scheduling
subsystem. In particular, increasing $K$
changes the priority resolution rather than
the number of parallel VN processing units.

\subsection{FPGA and ASIC Synthesis Results}
\label{subsec:synthesis}

The five decoder architectures were described
in synthesizable Verilog RTL and evaluated
using FPGA and ASIC synthesis flows.
FPGA resource utilization was obtained using
Xilinx Vivado, while ASIC synthesis was
performed using a 45-nm standard-cell
technology library. The evaluation focuses
on hardware resource consumption, silicon
area, and estimated power.

\textit{1) FPGA Synthesis:}
Table~\ref{tab:fpga_results} summarizes the
FPGA synthesis results. Conventional BP
requires 100,267 LUTs and 71,995 flip-flops
(FFs), reflecting the substantial hardware
resources associated with its flooding
implementation. RL-S reduces these values
to 69,737 LUTs and 10,536 FFs.
Further reductions are achieved by BRL-S,
which requires between 19,873 and 22,078
LUTs and approximately 5,800 FFs.

Compared with RL-S, the BRL-S implementations
reduce LUT utilization by 68.3--71.5\%
and FF utilization by 44.3--45.1\%.
The BRL2 configuration achieves the lowest
LUT count among the evaluated architectures,
while all three BRL-S configurations exhibit
similar FF utilization. No block RAM
(BRAM) or DSP resources are inferred
in the reported synthesis results. These results represent post-synthesis resource estimates rather than completed FPGA implementations.

\textit{2) ASIC Synthesis:}
The RTL designs were also synthesized
using a 45-nm standard cell library. Power consumption was estimated under identical operating conditions. We estimated the clock period using the slack from a 50ns clock constraint. Table~\ref{tab:asic_comparison} presents
the synthesized cell area and estimated
power consumption. The conventional BP
decoder occupies 1.0710~mm$^2$ and
consumes an estimated 15.243~mW,
whereas RL-S reduces these values to
0.4564~mm$^2$ and 3.766~mW. The BRL-S architectures achieve further
reductions in both metrics. BRL2 occupies
0.1485~mm$^2$ with an estimated power
consumption of 1.175~mW. BRL4 and BRL8
occupy 0.1591 and 0.1712~mm$^2$,
respectively. Relative to RL-S,
the three BRL-S implementations reduce
cell area by 62.5--67.5\% and estimated
power by 63.7--68.8\%.


Among the evaluated configurations, BRL2 achieves the lowest hardware cost, whereas BRL4 and BRL8 provide decoding performance closer to RL-S with modest additional resource consumption. These results demonstrate a favorable trade-off between error correction performance and hardware efficiency through priority quantization.


\section{Conclusion}
\label{sec:conclusion}

This paper presented hardware implementations of
conventional BP, reinforcement-learned sequential
BP (RL-S), and bucket-based reinforcement-learned
BP (BRL-S) for the $[[288,12,18]]$ bivariate bicycle
QLDPC code. By replacing max-heap priority scheduling
with quantized-priority FIFO buckets, BRL-S substantially
simplifies the scheduling hardware while retaining
the sequential message-update framework of RL-S.
Three BRL-S configurations with two, four, and eight
priority levels were evaluated through FPGA and
45-nm ASIC synthesis. Compared with RL-S, BRL-S
reduced FPGA LUT utilization by 68.3--71.5\%,
ASIC cell area by 62.5--67.5\%, and estimated
power consumption by 63.7--68.8\%.
Among the evaluated configurations, BRL2 achieved
the lowest hardware cost. These results demonstrate
that bucket-based scheduling provides a
hardware-efficient approach to implementing
reinforcement-learned QLDPC decoders.

\balance
\section*{Acknowledgment}

This work was supported by the
U.S. National Science Foundation under Grant
Nos. 2601103 and 2145917.

OpenAI ChatGPT was used to assist in drafting and
editing parts of the Introduction, Background, Hardware
Implementation, and Conclusion sections, and in
preparing Fig.~1. All technical content, equations,
and results were reviewed and verified by the authors.



\vspace{12pt}


\begin{thebibliography}{00}

\bibitem{QLDPC}
P. Panteleev and G. Kalachev,
``Asymptotically good quantum and locally testable
classical LDPC codes,'' \emph{Proc. 54th Annu. ACM SIGACT Symp.
Theory Comput. (STOC)}, pp. 375--388, 2022.

\bibitem{BBCode}
S. Bravyi, A. W. Cross, J. M. Gambetta,
D. Maslov, P. Rall, and T. J. Yoder,
``High-threshold and low-overhead fault-tolerant
quantum memory,''
\emph{Nature}, vol. 627, no. 8005,
pp. 778--782, Mar. 2024.

\bibitem{valls2021syndrome}
J. Valls, F. Garcia-Herrero, N. Raveendran, and B. Vasi{\'c},
``Syndrome-based min-sum vs. OSD-0 decoders:
FPGA implementation and analysis for quantum LDPC codes,''
\emph{IEEE Access}, vol. 9, pp. 138734--138743, 2021.

\bibitem{raveendran2023soft}
N. Raveendran, J. Valls, A. K. Pradhan, N. Rengaswamy,
F. Garcia-Herrero, and B. Vasi{\'c},
``Soft syndrome iterative decoding of quantum LDPC codes
and hardware architectures,''
\emph{EPJ Quantum Technol.}, vol. 10, Art. no. 45, 2023.

\bibitem{QuantumBP}
D. Poulin and Y. Chung,
``On the iterative decoding of sparse quantum codes,''
\emph{Quantum Information and Computation},
vol. 8, no. 10, pp. 987--1000, Nov. 2008.

\bibitem{ducrest2024checkagnosia}
J. du Crest, F. Garcia-Herrero, M. Mhalla,
V. Savin, and J. Valls,
``Check-agnosia based post-processor for message-passing
decoding of quantum LDPC codes,''
\emph{Quantum}, vol. 8, Art. no. 1334, May 2024.

\bibitem{bascones2025hardware}
D. B{\'a}scones, F. Garcia-Herrero, and J. Valls,
``Exploring the FPGA and ASIC design space of belief
propagation and ordered statistics decoders for quantum
error correction codes,''
\emph{EPJ Quantum Technol.}, vol. 12, Art. no. 140,
Nov. 2025.

\bibitem{zhou2025vegapunk}
K. Zhou, L. Lu, D. Xiang, C. Tao, A. Wu,
J. Leng, F. Liu, M. Chen, and J. Yin,
``Vegapunk: Accurate and fast decoding for quantum
LDPC codes with online hierarchical algorithm and
sparse accelerator,'' \emph{Proc. 58th IEEE/ACM Int. Symp.
Microarchitecture (MICRO)}, pp. 719--732,
2025.

\bibitem{maan2026correlated}
A. S. Maan, F. M. Garcia Herrero, A. Paler,
and V. Savin,
``Decoding correlated errors in quantum LDPC codes,''
\emph{Nature Commun.}, vol. 17, Art. no. 3965,
Mar. 2026.

\bibitem{Moradi2026Sequential}
M. Moradi, S. Habib, V. Nourozi,
and D. G. M. Mitchell,
``Sequential BP-based decoding of QLDPC codes,''
in \emph{Proc. IEEE Int. Conf. Commun. (ICC)}, pp. 1--6,
2026.

\bibitem{RLS}
M. Moradi, V. Nourozi, S. Habib,
and D. G. M. Mitchell,
``Learning to decode quantum LDPC codes via
belief propagation,''
arXiv preprint arXiv:2603.10192, Mar. 2026.

\bibitem{BRL}
M. Moradi, G. Aissou, T. Kim, R. A. Chou,
and D. G. M. Mitchell,
``Low-latency belief-propagation decoding of
quantum LDPC codes with reinforcement-learned
scheduling,''
arXiv preprint arXiv:2610.15857, Oct. 2026.

\bibitem{CalderbankShor}
A. R. Calderbank and P. W. Shor,
``Good quantum error-correcting codes exist,''
\emph{Physical Review A}, vol. 54, no. 2,
pp. 1098--1105, Aug. 1996.

\bibitem{Steane}
A. M. Steane,
``Multiple-particle interference and quantum
error correction,''
\emph{Proceedings of the Royal Society of London A},
vol. 452, no. 1954, pp. 2551--2577, Nov. 1996.

\bibitem{Andrade2015}
J. Andrade, N. George, K. Karras, D. Novo,
V. Silva, P. Ienne, and G. Falcao,
``From low-architectural expertise up to
high-throughput non-binary LDPC decoders:
Optimization guidelines using high-level synthesis,''
\emph{Proc. 25th Int. Conf. Field Programmable
Logic and Applications (FPL)}, pp. 1--8, 2015.

\bibitem{Ye2025ISCAS}
B. Ye, J. Liu, and H. Li,
``FPGA accelerated adaptive LDPC-based quantum
error correction by bitwise pipeline parallelism,'' \emph{Proc. IEEE Int. Symp. Circuits and
Systems (ISCAS)}, 2025.

\end{thebibliography}
\end{document}